\documentclass[aps,prd,twocolumn,floatfix,nofootinbib]{revtex4}
\usepackage{graphicx} \usepackage{epsf}
 
\def\gsim{ \lower .75ex \hbox{$\sim$} \llap{\raise .27ex \hbox{$>$}} }
\def\lsim{ \lower .75ex \hbox{$\sim$} \llap{\raise .27ex \hbox{$<$}} }
\def\be{\begin{equation}}  \def\ee{\end{equation}}

\def\mgut{m_{\mbox{\tiny GUT}}}
 
\begin{document}

\title{Boundary Effective Field Theory and Trans-Planckian Perturbations: \\Astrophysical Implications }

\author{Richard Easther$^{1}$}  \author{William H. Kinney$^{2}$}
\author{Hiranya Peiris$^3$\footnote{Hubble Fellow}}
 
\affiliation{~} \affiliation{$^1$Department of Physics, Yale
University, New Haven  CT 06520, USA \\ Email: {\tt
richard.easther@yale.edu}} 
 \affiliation{~}
\affiliation{$^2$Dept.\ of
Physics, University at Buffalo, SUNY, Buffalo, NY 14260, USA \\ Email:
{\tt whkinney@buffalo.edu}} 
 \affiliation{~}
\affiliation{$^3$Kavli Institute for Cosmological Physics and Enrico Fermi Institute,
University of Chicago, Chicago IL 60637, USA \\ Email: {\tt
hiranya@cfcp.uchicago.edu}}
 
\begin{abstract} 
We contrast two approaches to calculating trans-Planckian corrections to the inflationary perturbation spectrum: the New Physics Hypersurface [NPH] model, in which modes are normalized when their physical wavelength first exceeds a critical value, and the Boundary Effective Field Theory [BEFT] approach, where the initial conditions for all modes are set at the same time, and modified by higher dimensional operators enumerated via an effective field theory calculation. We show that these two approaches -- as currently implemented -- lead to radically different expectations for the trans-Planckian corrections to the CMB and emphasize that in the BEFT formalism we expect the perturbation spectrum to be dominated by quantum gravity corrections for all scales shorter than some critical value. Conversely, in the NPH case the quantum effects only dominate the longest modes that are typically much larger than the present horizon size. Furthermore,  the onset of the breakdown in the standard inflationary perturbation calculation predicted by the BEFT formalism
is likely to be associated with a  feature in the perturbation spectrum, and we discuss the  
observational signatures of this feature in both CMB and large scale structure observations. Finally, we discuss possible modifications to both calculational frameworks that would resolve the contradictions identified here. 
\end{abstract} 
 
\maketitle

\section{Introduction}

The possibility  that ``trans-Planckian'' effects modify the conventional understanding of inflationary perturbations is one of the most active topics in very early universe cosmology.  However, after several years' work there is still no theoretical consensus regarding the status of quantum gravitational or stringy modifications to the inflationary perturbation spectrum   \cite{Brandenberger:1999sw}-\cite{Greene:2005aj}.   Many possible models of trans-Planckian physics that modify the primordial perturbation spectrum have been examined, but none  amount to an {\it ab initio\/} calculation in a well-motivated ultraviolet completion of Einstein gravity, such as a string theory.

On dimensional grounds, we expect that the impact of ``new physics'' on the inflationary  perturbations depends upon $(H/M)^p$ where $H$ is the inflationary Hubble scale, $M$ is the scale at which the new physics comes into play, and $p$ depends on the  specific model of trans-Planckian physics.    We expect that $M$ will be within one or two orders of the Planck scale. The inflationary scale is constrained from above by the observational  bound on tensor contributions to the Cosmic Microwave Background [CMB], but roughly speaking $H$ is at or below the GUT scale.  In some inflationary models $H \ll \mgut$ and $H/M$ is thus tiny, independently of the value of $p$.  Even if $H \sim \mgut$, we need $p \lesssim1$ if it we are to disentangle any effect from uncertainties due to cosmic variance.  

We can   foresee three possible outcomes to the trans-Planckian problem. 
\begin{enumerate}
\item Quantum gravitational effects could completely upend our understanding of the generation of inflationary perturbations, and the primordial power spectrum can only be reliably computed after quantum gravitational effects are properly taken into account. 
\item A thorough calculation will show  trans-Planckian physics has no detectable impact on the inflationary spectrum, either because $p$ is large or $H/M$ is very small. 
\item Trans-Planckian effects modify the standard inflationary calculation, but they are always subdominant.  
\end{enumerate}
Obviously any of these options would have immediate consequences for cosmology.  The second is the most conservative, but it would be extremely reassuring to know that the standard inflationary perturbation was not modified by quantum gravitational effects, especially as we are entering an era in which observational tests of inflationary models are increasingly precise.  The third option is perhaps the most immediately interesting, 
 since it preserves one of the main predictions of inflation -- an approximately scale-free spectrum of primordial perturbations -- but still holds out the hope that astrophysical observations will one day permit string theory and/or quantum gravity to be tested in ways that are completely inaccessible to any imaginable terrestrial experiments. Finally, from a theoretical perspective the first option would be the most fascinating since it would imply that observational cosmology can directly test the stringy / quantum gravity regime but at the cost of being able to obtain a detailed prediction for the spectrum using presently available theoretical tools.

Arguably, the most promising theoretical approach to the trans-Planckian problem that has been tried to date is the {\em Boundary Effective Field Theory\/} [BEFT] formalism developed by Schalm, Shiu, van der Schaar \cite{Schalm:2004qk,Schalm:2004xg} and developed further   in collaboration with Greene \cite{Greene:2004np,
Greene:2005aj}. 
   Here, the machinery of effective field theory is used to identify all the possible higher-dimension operators that could modify either the initial state or subsequent evolution of the quantum mechanical fluctuations that grow into cosmological perturbations.  The strength of the BEFT approach is that it formulates the problem in
a well-controlled field theoretic context, and allows a self-consistent
investigation of issues such as the back reaction of ultraviolet modes.   In this picture, the details of the quantum mechanical corrections to ``standard'' gravity are encoded in the coefficients of these operators, and the lowest order operator is found to modify the initial state of the perturbations by an amount proportional to $H/M$ -- corresponding to $p=1$ and leaving open the possibility that the trans-Planckian physics may have observationally accessible consequences.   

Taking the spectrum derived by Schalm {\it et al.\/} at face value, we will see that it predicts that either the cosmological perturbations are effectively identical to the standard result, or that the corrections from trans-Planckian effects are of the same magnitude as the spectrum itself. In the latter case we would therefore need  a full quantum gravitational calculation to make a reliable prediction for the spectrum.  There is a small range of intermediate scales where the corrections associated with the minimum length are small but non-zero.   Consequently, the BEFT formalism appears to imply that instead of a small correction to the usual inflationary spectrum (Option 3, the outcome of many other approaches to the trans-Planckian corrections to the perturbation spectrum), we can expect either a dramatic modification to the spectrum or no effect at all (Options 1 or 2).   In Schalm {\em et al.}'s calculation the scale at which the standard inflationary result breaks down is an arbitrary parameter, although it would presumably be predicted by an {\em ab initio\/} calculation.     If this breakdown occurs at astrophysically interesting scales quantum gravity effects are dominant at all {\em shorter\/} scales, which will necessarily be inside our cosmological horizon. However, given that this form of the  BEFT spectrum  implies a radical breaking of the approximate time invariance of an inflating universe it is premature to conclude that we are faced with a choice between Options 1 and 2.  Rather, this is a signal that  the BEFT approach as it stands is incomplete, and provides additional motivation for constructing a BEFT  formalism where the initial conditions are set on a momentum hyperplane, as is currently done in the New Physics Hypersurface [NPH] formalism  \cite{Danielsson:2002kx,Easther:2002xe}.  Also, note that with the benefit of hindsight,  we can see that \cite{Starobinsky:2001kn} writes down a set of initial conditions corresponding to the choice made in the NPH analysis. 
 
A number of studies have addressed the observability of trans-Planckian corrections to the perturbation spectrum \cite{Bergstrom:2002yd,Elgaroy:2003gq,Okamoto:2003wk,Martin:2003sg,Martin:2004yi,Easther:2004vq}. This paper extends our recent analysis of the possible cosmological constraints that can be placed on the NPH formalism \cite{Easther:2004vq}, as we explore the cosmological consequences of the BEFT formalism, in order to  better understand its astrophysical implications.      In Section \ref{sec:spectra} we summarize the predictions of  the BEFT formalism, contrast  the results of calculations based on the NPH approach, and discuss the role of the free parameters in the BEFT formalism.  In Section \ref{sec:matching} we look more carefully at the assumptions that go into setting the parameters for the BEFT model, and show that it is entirely possible (although not guaranteed) that the BEFT model makes no detectable difference to  the standard inflationary perturbation spectrum.  In Section \ref{sec:impact} we briefly consider the possible impact of the BEFT-modified spectrum on the CMB and on the formation of large scale structure. In Section \ref{sec:runningbeta}  we look at possible ways in which the two approaches might be reconciled,  and discuss whether the BEFT formalism is complete as it currently stands. We conclude in Section \ref{sec:conclusion}.
 
 \section{The BEFT and NPH Spectra}
 \label{sec:spectra}

In the BEFT formalism,  the spectrum is related to the usual Bunch-Davies result ($P_{BD}$) as follows \cite{Greene:2005aj} :
\begin{eqnarray} 
\label{eq:EFTmod} 
P_{BEFT}(k) & = & P_{BD}(k) \left[ 1 + {\beta  k \over a_i M} \sin\left(2  { k \over a_i H_i}\right)\right] \, ,  \\
                     &=&  P_{BD}(k) \left[ 1 + \beta {H_i \over M} y_i \sin\left(2 y_i\right)\right].
\end{eqnarray}
Here $\beta$ is an unknown coefficient of order unity that appears in front of the leading-order irrelevant operator in the effective field theory computation. The scale of new physics is set by  $M$ which is assumed to be within a couple of orders of magnitude of the Planck scale, $k$ is the comoving wavenumber, and $H_i$ is the value of the Hubble constant during inflation.   Finally, $a_i$ is the value of the scale factor upon the hypersurface where  all the initial conditions are set in the BEFT approach. Here $y = k/aH$ is a rescaled ``time'' variable that decreases from the large positive values towards zero during inflation, where the moment of horizon exit is (by definition) $y=1$ \cite{Kinney:1997ne}.   Compare this with the result obtained from the NPH formalism \cite{Danielsson:2002kx,Easther:2002xe},   
\begin{equation} 
\label{eq:PNPH}
P_{NPH}(k) = P_{BD}(k) \left[ 1 +  {1 \over 2 y_c} \sin\left(  {2 y_c  \over 1-\epsilon} \right)\right] \, ,
\end{equation}
where $y_c$ here is the value of $y$ when the physical wavelength of the mode first exceeds the minimum length $1/M$ and $\epsilon$ is the slow roll parameter, which goes to zero in the de Sitter limit.   In this case $y_c \sim k^\epsilon$ is weakly scale dependent, since $\epsilon \ll 1$ for realistic inflationary models.       Note that in the spectra above we have suppressed an arbitrary phase that arises due to our ignorance of the detailed post-inflationary evolution and the consequent absence of a precise result for the moment of horizon exit during the inflationary era for a given comoving scale.  

When we compare the two cases, we see  that while both formalisms lead to a modulated spectrum, the details differ greatly.  For the NPH model, the modulation decreases with increasing $k$, so we see a smaller deviation at short scales relative to that seen at longer ones. Moreover,  since $y_c \sim (k/k_\star)^\epsilon \sim e^{\epsilon \log{k/k_\star}} \sim 1 + \epsilon \log{k/k_\star}$ we see that the modulation is {\em logarithmic\/} in  $k$.   In the BEFT case, conversely, the amplitude of the modulation is growing linearly with  $k$. Moreover, we see that the phase of the  modulation is proportional to $k$, and is thus potentially much more rapid than in the NPH case, where the modulation if effectively a function of $\log{k}$. 

\begin{figure} 
\includegraphics[width=3in]{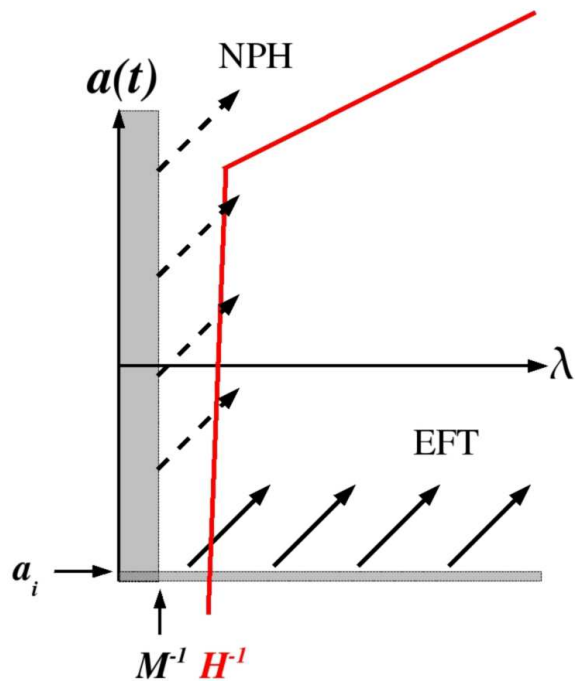}
\caption[]{\label{fig:spacetime}Schematic diagram of the initial conditions hypersurfaces for the BEFT and NPH models of trans-Planckian physics.  For the NPH case, the  initial conditions are set when the wavelength $a/k$ is equal to the minimum length $1/M$, and all modes are normalized on a vertical timelike hypersurface. Conversely, in the BEFT case the initial conditions for  all the modes are set on the same spacelike hypersurface, where $a=a_i$.  The physical wavelength of a given mode is measured by $\lambda$.}
\end{figure}

The physical difference between the two formalisms is shown schematically in Figure~(\ref{fig:spacetime}).  In the NPH formalism, each mode is normalized when its physical wavelength is equal to $1/M$. During inflation, $H/M$ changes slowly, and so two modes with similar $k$ will share a similar normalization.   In the BEFT case all modes are normalized on the same spacelike hypersurface.  Here long modes are {\em already\/} outside the horizon when $a=a_i$ and the BEFT calculation is understood as encoding all their previous evolution.  Conversely, the short wavelength modes are normalized when they lie well inside the horizon and, in most cases, when their physical wavelength is much shorter than the minimum length $1/M$.   It is these modes, which have enough momentum on the initial hypersurface to directly sample the trans-Planckian physics, for which the BEFT spectrum is not reliable.

\begin{figure*}
\includegraphics[width=16cm]{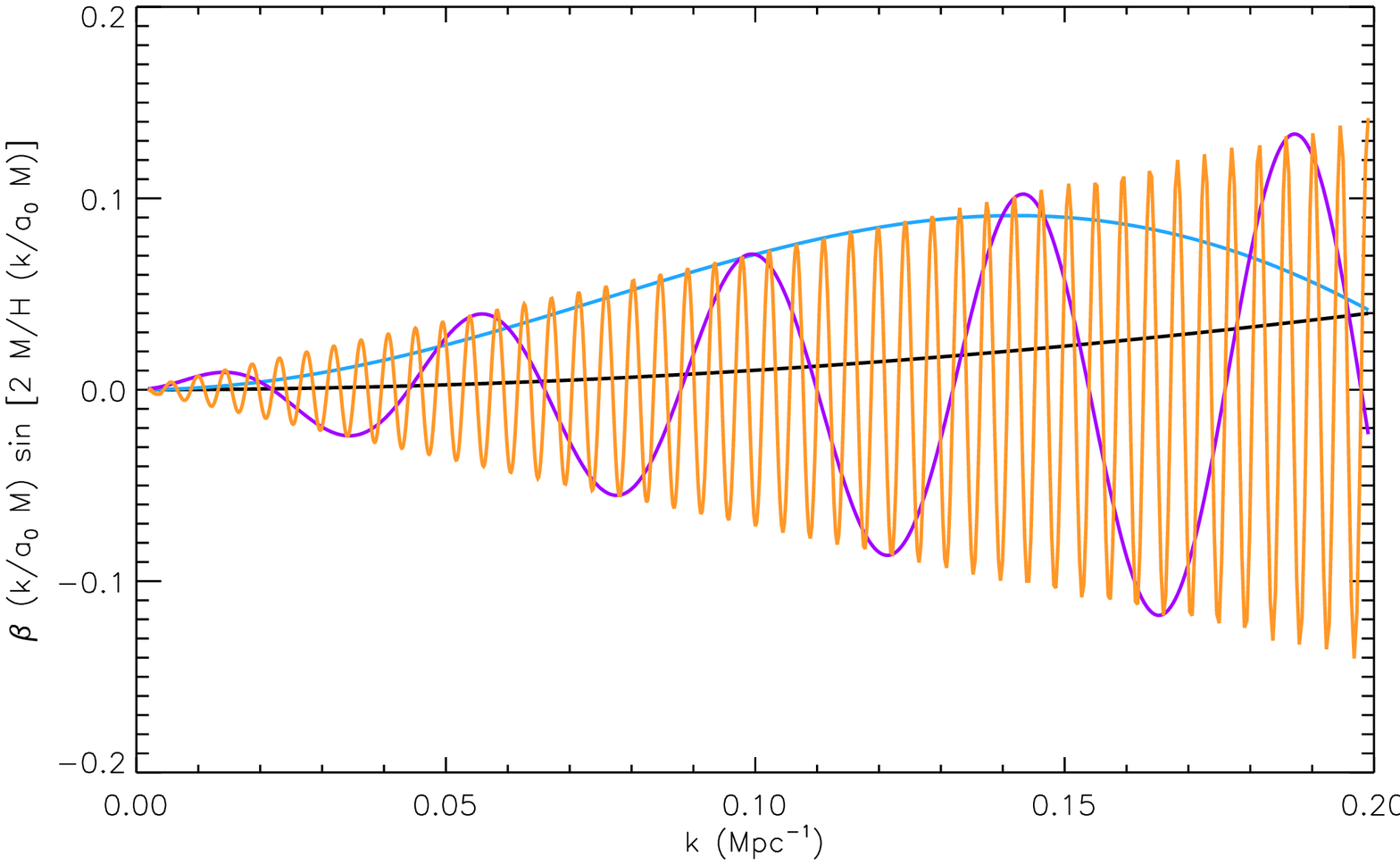}
\caption[]{\label{fig:PSmod} Modification to the primordial power spectrum for four choices of parameters for the Boundary Effective Field Theory, chosen so that the modulation has an impact on the CMB. The plots show $\beta=1$, $a_i M = 1.4\ \mbox{Mpc}^{-1}$  and $M/H = 1$ (black), 10 (blue), 100 (purple), 1000 (orange).}
\includegraphics[width=16cm]{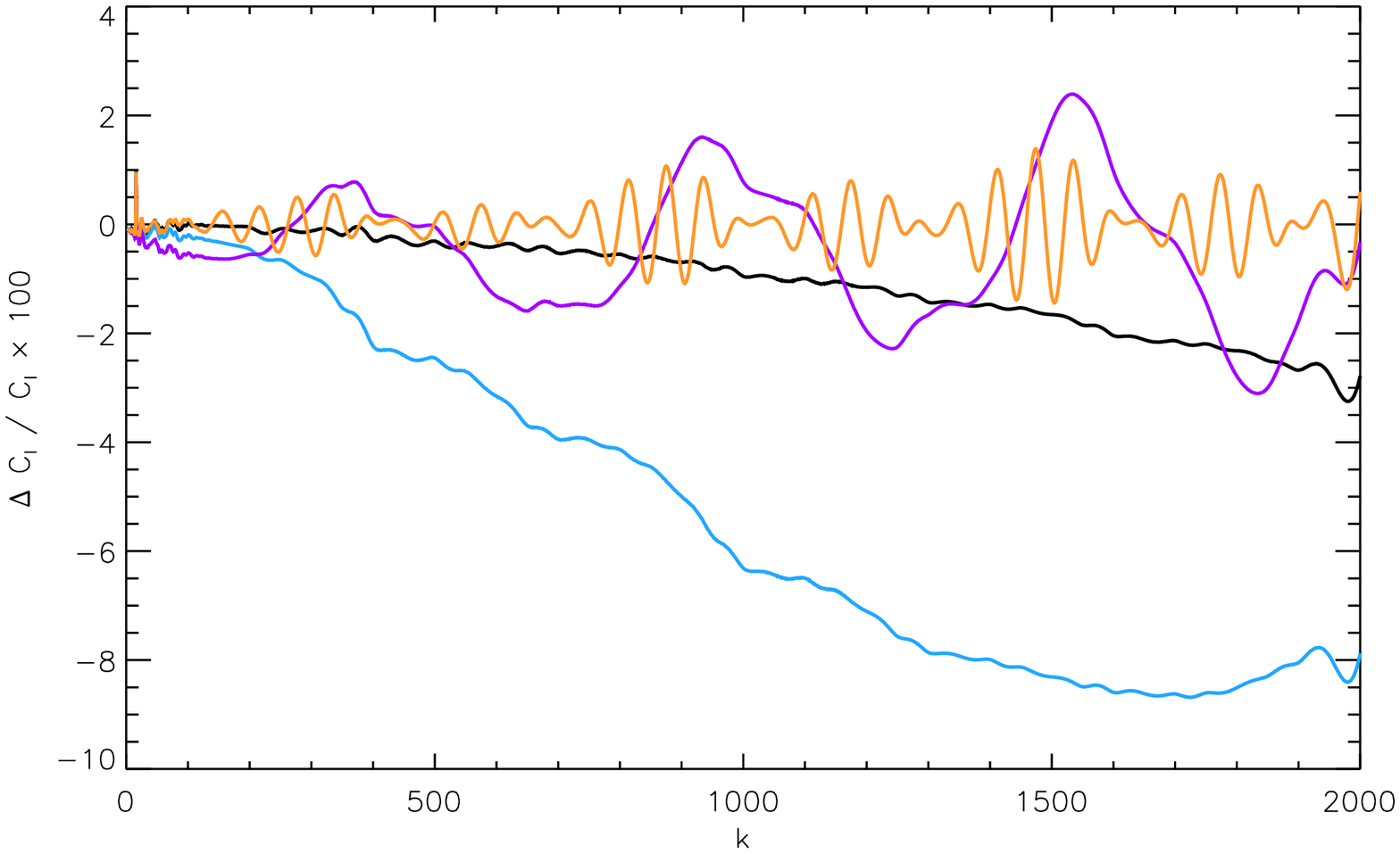}
\caption[]{\label{fig:clmod} Modification to the CMB multipole spectrum for four choices of parameters for the Boundary Effective Field Theory. Parameter choices match those of Figure \ref{fig:PSmod}.}
\end{figure*}

From an observational standpoint, the key difference between the two spectra is that in the NPH case the modes approach the Bunch-Davies vacuum at short wavelengths, whereas in the BEFT case the shorter modes depart further from the Bunch-Davies case than the long ones.  In both models, the modulation is of order unity for some critical value of $k$, which we associate with a breakdown of the formalism used to do the calculation.  In the NPH case, this corresponds to the minimal length being roughly equal to the horizon size. For the BEFT case,  the amplitude of the correction grows with $k$, and while we may have a reliable result at large scales we have no prediction at  all for the spectrum when $y_i > M/H_i \beta$. If the critical value of $k$ corresponds to a very short wavelength (relative to astrophysical scales) then there will be a negligible correction at long scales.  On the other hand, if the critical value of $k$ is small enough to influence perturbations that contribute to the microwave background, then we have no prediction at all for $P(k)$ for modes that contribute to structure formation or the Lyman-$\alpha$ forest.  Four representative BEFT modified spectra are shown in Figure~(\ref{fig:PSmod}).  The spectrum is always oscillatory, but the oscillation is faster for models where $M \gg H$.

\section{Matching to Astrophysical Scales}
\label{sec:matching}

\begin{figure*}
\includegraphics[width=16cm]{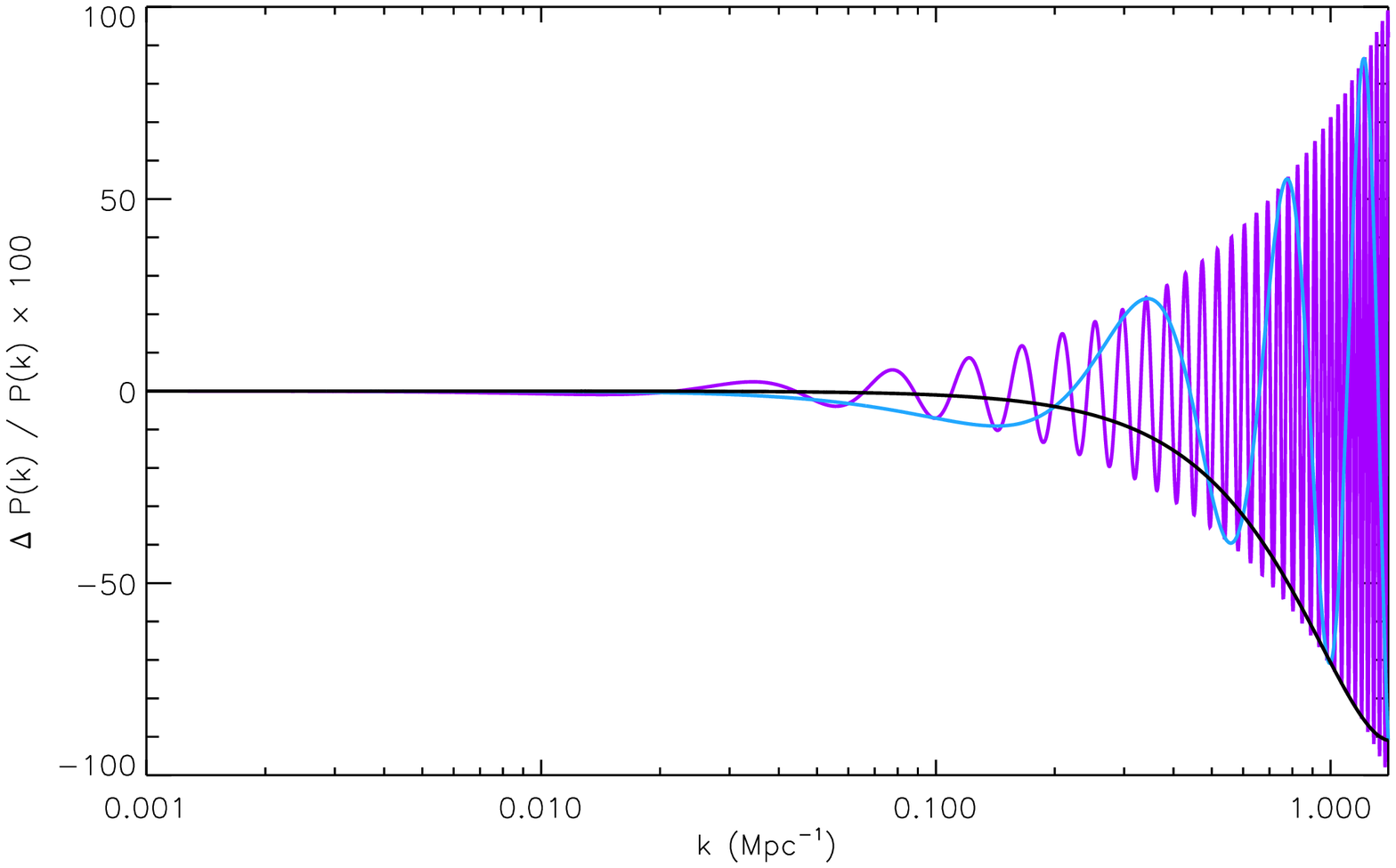}
\caption[]{\label{fig:PKmod} Modification to the linear theory matter power spectrum for three choices of parameters for the Boundary Effective Field Theory. The plots show $\beta=1$, $a_i M = 1.4\ \mbox{Mpc}^{-1}$  and $M/H = 1$ (black), 10 (blue), and 100 (purple). The single decade of $k$ values for which the modulation is measurable and where the BEFT spectrum is reliable is $0.01 < k < 0.1$.}
\end{figure*}

For a given choice of boundary hypersurface in the BEFT formalism, the frequency of the modulation (\ref{eq:EFTmod}) is determined by the size of the mode relative to the horizon on the boundary hypersurface, and the amplitude of the modulation is determined by the size of the mode relative to the cutoff scale $M$. For modes of order the cutoff scale on the boundary hypersurface, we expect the modulation in the power spectrum to be of order unity for $\beta \sim O(1)$. Note in particular that since the power spectrum is positive-definite, once the coefficient $k / (a_i M)$ becomes larger than one, this expression for the modulation breaks down. This is consistent with the expectation in effective field theory that we are not allowed to consider modes with wavelength shorter than the cutoff. So the modulation of the power spectrum in the BEFT formalism is largest for the shortest wavelengths ({\it i.e.} the modes closest to the cutoff scale on the boundary hypersurface) and becomes small in the long-wavelength limit. This behavior is very different from the NPH case. In the NPH case, the modulation is largest for long-wavelength modes, {\it i.e.} modes which left the horizon earlier in inflation when the horizon size was closer to the cutoff scale. 

There is an inherent ambiguity in the BEFT prescription, corresponding to the freedom to choose the initial hypersurface, or, equivalently, $a_i$. Naively, one might expect that any physically measurable result like the modulation of the power spectrum should be independent of our choice of $a_i$, but the expression (\ref{eq:EFTmod})
manifestly depends on  $a_i$. One possible way the dependence of the modulation on the choice of initial hypersurface might be modified is if different choices of $a_i$ result in a renormalization of the effective coupling $\beta$. In what follows, we will assume that $\beta$ is an order-unity constant whose value does {\em not} depend on the choice of $a_i$. With this assumption, the amplitude and frequency of the power spectrum modulation which we expect to observe in the CMB are strongly dependent on our choice of $a_i$. We will return to the possibility of a running value of $\beta$ in Section \ref{sec:runningbeta}.

For a given choice of initial hypersurface $a_i$, it is easy to calculate the physical size of a scale today in terms of its size on the boundary hypersurface,
\begin{equation}
\label{eq:scaleratio1}
{y_{\rm today} \over y_i} = {(a H)_{\rm today} \over (a H)_i},
\end{equation}
where $y \equiv k / (a H)$ is the size of a mode $k$ relative to the horizon.
The CMB quadrupole corresponds to a scale of order the current horizon size, $y_{\rm today} \sim 1$.
The current scale factor $a_{\rm today}$ can be written in terms of the scale factor at the end of inflation $a_e$,
\begin{equation}
{a_e \over a_{\rm today}} = {T_{\rm RH} \over T_{\rm CMB}} \sim {\sqrt{H_i m_{\rm Pl}} \over T_{\rm CMB}},
\end{equation}
where $T_{\rm CMB} = 2.7\ K$ is the current CMB temperature, and the reheat temperature is $T_{\rm RH} \sim \sqrt{H_i m_{\rm Pl}}$, assuming instantaneous reheating. We can write $a_e$ in terms of the the scale factor $a_i$ on the initial hypersurface using the number of e-folds before the end of inflation,
\begin{equation}
{a_e \over a_i} = e^{N_i}.
\end{equation}
Note that we use the usual convention of $N = 0$ at the end of inflation, and growing as we go backward in time. We have just rewritten our freedom to choose $a_i$ in terms of the parameter $N_i$. Then we have
\begin{equation}
{a_{\rm today} \over a_i} = e^{N_i} \left({\sqrt{m_{\rm Pl} H_i} \over T_{\rm CMB}}\right).
\end{equation}
Then the expression (\ref{eq:scaleratio1}) becomes
\begin{equation}
{y_{\rm today} \over y_i} \sim e^{-N_i} \left({T_{\rm CMB} \over H_{\rm today}}\right) \sqrt{H_i \over m_{\rm Pl}} \sim 10^{29} e^{-N_i} \sqrt{H_i \over m_{\rm Pl}}.
\end{equation}
Here $(H_i / m_{\rm Pl}) \sim 10^{-4}$ is the ratio of the Hubble parameter during inflation to the Planck scale. 
Therefore the scale corresponding to the CMB quadrupole is given by a scale on the initial hypersurface of order
\begin{equation}
y^{\rm quad}_i \sim 10^{-29} e^{N_i} \sqrt{m_{\rm Pl} \over H_i}.
\end{equation}
Scales of order the horizon size today crossed out of the the horizon during inflation when $y^{\rm quad}_i \sim 1$, or $N_i \sim 62$, assuming $H_i \sim \mgut^2/  m_{\rm Pl}$.

The question of a physically sensible choice of boundary hypersurface then translates into a physically sensible choice of $N_i$. If we wish the BEFT calculation to be predictive on all astrophysical scales, this corresponds to a choice of boundary surface such that {\em any mode which exits the horizon during inflation has wavelength larger than the cutoff on the boundary surface}.  Note that observations of close quasar pairs may probe the Lyman-$\alpha$ forest down to 25 Mpc$^{-1}$ \cite{Lidz:2003fv} and if we can directly observe the stochastic gravitational wave background produced during inflation, we will be probing inflationary perturbations on scales corresponding to the last few e-folds of inflation. That is, the fluctuation mode which just exits the horizon at the end of inflation must be no smaller than the cutoff on the boundary surface, which corresponds to a condition on $N_i$ of
\begin{equation}
e^{N_i} \sim {M \over H_i}.
\end{equation}
Then the quadrupole scale on the boundary is:
\begin{equation}
y^{\rm quad}_i \sim 10^{-29} {\sqrt{M m_{\rm Pl}} \over H_i},
\end{equation}
and the amplitude of the modulation at the quadrupole is of order
\begin{equation}
\left({\delta P(k) \over P(k)}\right)_{\rm quad} \sim {H_i \over M} y^{\rm quad}_i \sim 10^{-29} \sqrt{m_{\rm Pl} \over M}.
\end{equation}
If we select $a_i$ such that all fluctuations which exit the horizon during inflation have a power spectrum which is well-defined in the BEFT, we have an exponentially suppressed modulation on CMB scales. We are, however, free to choose our initial hypersurface earlier in inflation, {\it i.e.} at bigger $N_i$.  If we pick the initial hypersurface to be the time when the quadrupole exits the horizon, $N_i \sim 62$, the modulation of the power spectrum at the quadrupole is
\begin{equation}
\left({\delta P(k) \over P(k)}\right)_{\rm quad} \sim 10^{-2} {\sqrt{m_{\rm Pl} H_i} \over M}.
\end{equation}
If we take $M \sim m_{\rm Pl}$ and $H_i / m_{\rm Pl} \sim 10^{-4}$, the modulation at the quadrupole is of order $\delta P / P \sim 10^{-4}$ at the quadrupole, and grows to of order $\delta P / P \sim 0.1$ at around $\ell = 1000$ -- an effect which would be easily observable, as we shall see in the next section. Moving to shorter wavelengths, the modulation grows to of order unity at around $\ell = 10,000$, and the BEFT formalism breaks down entirely for length scales shorter than that. The boundary BEFT therefore makes no prediction at all for perturbations on scales corresponding to large-scale structure surveys or observations the Lyman-alpha 
forest.

\section{Astrophysical Impact of a BEFT-modified Spectrum}
\label{sec:impact}

\begin{figure*}
\includegraphics[width=16cm]{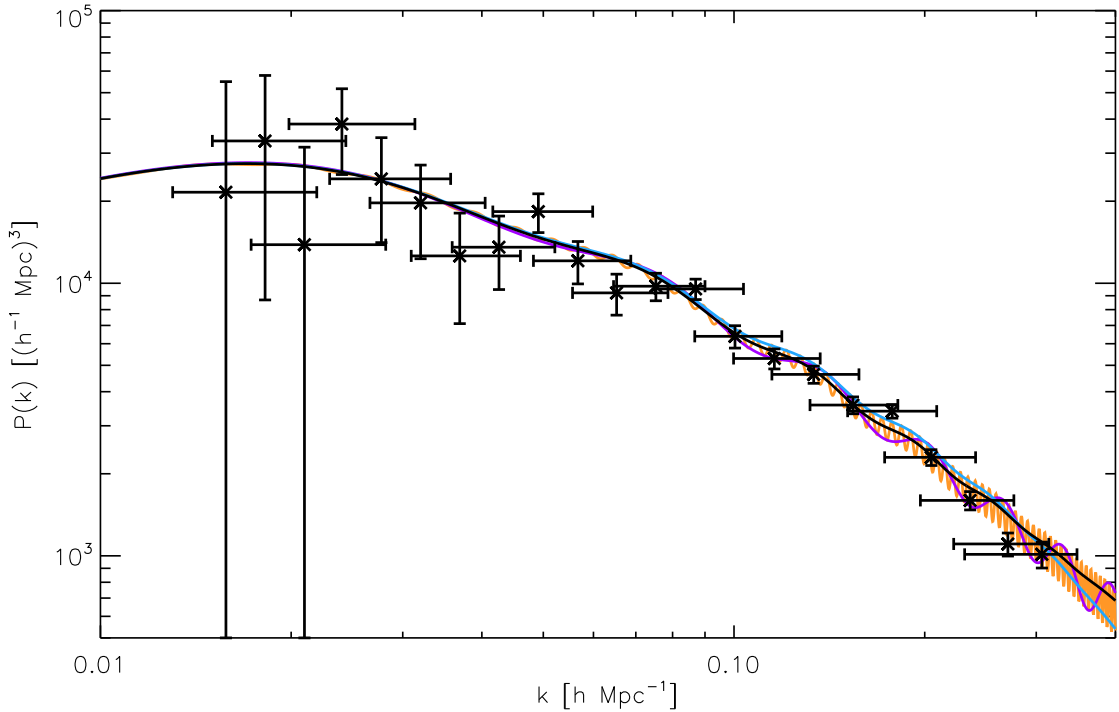}
\caption[]{\label{fig:pksdss} Modification to the linear theory matter power spectrum for four choices of parameters for the Boundary Effective Field Theory, plotted along with error bars from SDSS. Parameter choices match those of Figure \ref{fig:PSmod}. As a rough guide, WMAP \cite{Bennett:2003bz} probes $k < 0.06$ Mpc$^{-1}$, SDSS probes $0.015$ Mpc$^{-1} < k < 0.3$ Mpc$^{-1}$, and the Lyman-$\alpha$ power spectrum probes $0.2$ Mpc$^{-1} < k < 4$ Mpc$^{-1}$. Note the observations of the Lyman-$\alpha$  forest for close quasar pairs may probe scales as short as 25 Mpc$^{-1}$ \cite{Lidz:2003fv}.  
}
\end{figure*}

We have seen in the previous section that the boundary BEFT formalism does not make a self-consistent prediction of the form of the primordial power spectrum on scales smaller than the cutoff scale, $k > a_i M$. The BEFT cutoff is not a physically constant length scale, but a {\em comoving} scale, since we are integrating out modes with momentum larger than  $k_c = a_i M$, which is constant in comoving units. This does not mean that modes shorter than the cutoff are unphysical, but simply that they are absorbed into higher-order operators by our choice of expansion. This is a far more radical proposal than previous attempts to incorporate trans-Planckian physics into the inflationary perturbation framework, which almost universally assume a cutoff with a fixed physical, rather than comoving, scale. In the BEFT formalism, modes with {\em comoving} momentum higher than $k_c = a_i M$ will be strongly influenced by quantum-gravitational effects, and therefore cannot be calculated using a low-energy effective theory. If the comoving cutoff scale is large enough in current astrophysical units, we should be able to directly probe strongly quantum-gravitational effects by observing cosmological structure. (The cutoff is imposed on the boundary surface and therefore applies to the {\em initial state} and not the subsequent dynamics: the cutoff on the dynamics of the theory is constant in physical, not comoving units.)

The fact that existing structure surveys such as the 2dF Galaxy Redshift Survey \cite{Percival:2001hw, Cole:2005sx} and the Sloan Digital Sky Survey \cite{Tegmark:2003uf} are consistent with a scale-free primordial power spectrum should put strong constraints on the nature of such physics. However, without a theory of the extreme ultraviolet behavior or an implementation of the BEFT formalism that is fully compatible with an expanding universe, it is impossible to do more than speculate on the form such constraints might take.  It is worth noting that Harrison and Zeldovich \cite{Harrison:1969fb,Zeldovich:1972ij} originally argued for a scale-free spectrum on very general grounds, and this expectation might well survive in a fully quantum gravitational calculation.  However, the normalization and scale dependence of the spectrum predicted of the standard inflationary spectrum need not match the full quantum gravitational /  stringy calculation required at short scales in the BEFT formalism. Thus, taking the BEFT of equation~(\ref{eq:EFTmod}) at face values suggests that we would expect to see a break in the primordial spectrum around $k \sim a_i M$.

We can, however, self-consistently investigate the cosmological effect of the trans-Planckian BEFT on modes with momentum lower than the comoving cutoff, $k < a_i M$. The BEFT makes a definite prediction for the power spectrum modulation for such modes. Looking at equation~(\ref{eq:EFTmod}), the amplitude of the modulation rises linearly with $k$.  If we adopt the rule of thumb that a modulation of less than 1\% is unobservable, then there are only two decades in $k$ for which the modulation is big enough to be seen, but for which the spectrum is still positive definite. This is true for any values we assign for $\beta$, $H$, $M$  or $a_i$. Changing these will alter the value of $k$ at which the breakdown occurs, but does not alter the linear dependence of the amplitude of the modulation on $k$.   If this really is a perturbative expansion, we might guess that the next-order corrections will  be on the order of 1\% when the modulation is on the order of 10\%, so we really only have one decade in $k$ over which we have a reliable expression for the spectrum as well as a measurable difference from the usual  Bunch-Davies result.

Looking at the the spectra plotted in Figures \ref{fig:PSmod}--\ref{fig:PKmod}, the average power is unchanged, so  a very rapid oscillation is ``smoothed out'' when we look at its impact on any astrophysical process. However once the modulation of the spectrum is of the same order as the spectrum itself, there is no guarantee that this averaging will be preserved, as higher order effects must come into play. Consequently, as noted above the naive astrophysical expectation is that the spectrum will contain  a ``feature''  in $k$ space at $k=k_c$. For $k \lesssim k_c$ we can look at the impact of the BEFT modulation on astrophysical phenomena but for $k \gsim k_c$ we have no reliable prediction for the power spectrum. 

We now turn to the detectability of a BEFT-induced modulation via astrophysical probes of the primordial spectrum.  Since the BEFT formalism (like most other trans-Planckian proposals) modifies the primordial power spectrum and not the detailed physics of recombination and the subsequent generation of temperature anisotropies, it is straightforward to calculate the resulting anisotropies in the CMB generated by an underlying spectrum with the form of equation~(\ref{eq:EFTmod}).  Using a modified version of CAMB\footnote{CAMB: {\tt http://camb.info/}} we have plotted CMB power spectra for temperature anisotropies for four different choices of parameters, where the critical value of $k$ has been chosen to correspond to 1.4 Mpc$^{-1}$.  Figure \ref{fig:PSmod} shows the power spectrum modulation and Figure~\ref{fig:clmod} shows the corresponding modulation to the $C_\ell$'s.  Here we can see the averaging effect in action -- as the oscillation becomes more rapid, the impact on the $C_\ell$'s is reduced. Clearly, the influence of trans-Planckian effects on the CMB spectrum can be strikingly large for a comoving cutoff sufficiently close to scales probed by the CMB. However, for sufficiently rapid oscillations, {\it i.e.}, large enough $k / (a_i H)$, it is to be expected that logarithmic binning of the power spectrum in $k$-space will tend to average over multiple oscillations, and the oscillation will be unobservable. With this caveat, we can conclude that a generic feature of the BEFT is a large (between of order 10\% and of order unity) modulation in the power spectrum near the comoving cutoff scale. If the comoving cutoff scale is larger than the current scale of nonlinear structure (and perhaps much smaller than that), the modulation is likely to either be detected or definitively ruled out by near-future observations.   
 
If $k_c$ is at scales much shorter than those which can contribute to the CMB, it is still possible  to see the impact of the BEFT corrected spectrum in either large scale structure observations such as the Sloan Digital Sky Survey (SDSS) \cite{Tegmark:2003uf}, or in Lyman-$\alpha$ forest measurements \cite{Seljak:2004xh,Lidz:2003fv}. One of the most startling successes of the standard paradigm of structure formation is that the overall normalization of the  primordial power spectrum obtained through observations of structures that have undergone non-linear gravitational collapse such as galaxies match those obtained from the linear perturbations in the CMB and Lyman-$\alpha$ clouds. Figure \ref{fig:pksdss} shows four BEFT spectra with cutoff scales of 1.4 Mpc$^{-1}$ plotted along with the current errorbars from the SDSS. While these spectra are consistent with the current SDSS data, as shown in Figure \ref{fig:PKmod} they acquire order unity modifications to the matter power-spectrum around the cutoff scale. Therefore, for these models, a break in the power spectrum should be detectable in Lyman-$\alpha$ data. If the BEFT calculation is correct and if $k_c$ corresponds to any scale between a few hundred kiloparsecs and the present size of the visible universe, it is very likely  we will soon have compelling evidence for this in the form of a large feature in the power spectrum. Conversely if $k_c$ is very much larger than any of these scales, then the impact of trans-Planckian physics is forever unobservable via the primordial power spectrum. 
 
\section{Restoring Time Invariance}
\label{sec:runningbeta}

Perhaps the most surprising feature of the BEFT formalism is  that the modulation to the power spectrum depends strongly on the choice of initial hypersurface. In an effective field theory description of low-energy physics, it is typical to expect a dependence of the effective Lagrangian on the choice of cutoff scale $M$, but using the BEFT prescription in an expanding spacetime adds an additional dependence on {\em when} one chooses to apply the boundary condition.   Physically, this corresponds to the time invariance  of a de Sitter universe being badly broken where BEFT corrections are added to the fundamental perturbation spectrum.  By contrast, the NPH approach respects the time invariance of the original background -- in the de Sitter limit, the NPH spectrum renormalizes the exactly scale-free de Sitter spectrum by a small factor, but it does not induce any scale dependence.  The modulation seen in the NPH spectrum arises from the slowly changing value of $H$ -- a weak breaking of time invariance that is endemic to slow roll inflation.

It is thus of interest to investigate the consequences of broken time invariance in the BEFT formalism, and whether we can impose the condition that  the modulation is independent of the choice of initial hypersurface. Expressing the modulation for a given wavenumber $k$ in terms of $y_i \equiv k / (a H)_i$, we can write the modulation as
\begin{equation}
\label{eq:specmod2}
{\delta P \over P} = 1 + y_i \beta(y_i) \left({H_i \over M}\right) \sin\left(2 y_i\right),
\end{equation}
where we have allowed the constant $\beta$ to be a function of our choice of initial hypersurface $\beta = \beta(y_i)$. This is bound to be a very rough analysis, since this expression is lowest-order in what is formally an infinite series of irrelevant operators with independent coefficients $\beta$. However, it will give us a qualitative idea of the change in the power spectrum modulation resulting from the imposition of time-invariance on the effective field theory expansion. 
The simplest case is de Sitter space, where $H = {\rm const.}$. If the modulation for a given mode $k$ is independent of the initial hypersurface, then
\begin{equation}
{d \over d y_i} \left({\delta P \over P}\right) = \left({H \over M}\right) {d \over dy}\left[y_i \beta(y_i) \sin\left(2 y_i\right)\right] = 0, 
\end{equation}
with the trivial solution
\begin{equation}
\beta(y_i) = {\beta_0 \over y_i \sin\left(2 y_i\right)} \, .
\end{equation}
This expression becomes large for $\sin\left(2 y_i\right) \rightarrow 0$, indicating a breakdown of the effective field theory expansion which will be especially acute for modes near the cutoff $k = a M$, for which $y_i \gg 1$ and the power spectrum modulation is rapidly oscillating. However, we expect this expression for the running of $\beta$ to be valid for a range of modes near horizon crossing on the initial hypersurface, $y_i \lsim 1$, and
\begin{equation}
\beta(y_i) \sim {\beta_0 \over y_i^2}.
\end{equation}
With this running of $\beta$, the $k$-dependence of the modulation is suppressed, and the power spectrum simply receives a change in normalization,
\begin{equation}
{\delta P \over P} = 1 + \beta_0 \left({H \over M}\right).
\end{equation}
This can be generalized to the case of non-de Sitter evolution by using the relation \cite{Kinney:2005vj}
\begin{equation}
{d \over d y} = {1 \over y \left(1 - \epsilon\right)} {d \over d N},
\end{equation}
where $N$ is the number of e-folds until end of inflation, and 
\begin{equation}
\epsilon = {3 \over 2} \left(1 + {p \over \rho}\right)
\end{equation}
is the first slow roll parameter. We can then use the flow function relation \cite{Kinney:2002qn}
\begin{equation}
{d H \over d N} = H \epsilon
\end{equation}
to write
\begin{equation}
\label{eq:HoverMy}
\left({H_i \over M}\right) \propto y_i^{\epsilon / (1 - \epsilon)}.
\end{equation}
We can then write the modulation of the power spectrum as:
\begin{equation}
{\delta P \over P} = 1 +  \left({H \over M}\right)_{y = 1} y_i^{1 / (1 - \epsilon)} \beta(y_i) \sin\left(2 y_i\right),
\end{equation}
where $(H/M)_{y = 1}$ is the ratio of the cutoff scale to the horizon size evaluated at the time the mode with wavenumber $k$ exited the horizon, $y = k / (a H) = 1$. Demanding that this expression be independent of $y_i$ results in
\begin{equation}
{\delta P \over P}\left(k\right) = 1 +  \beta_0 \left({H \over M}\right)_{y = 1},
\end{equation}
which is now a $k$-dependent modulation, since $H / M$ varies with time. We can use the well-known result\footnote{Note that, despite appearances, Eqs. (\ref{eq:HoverMy}) and (\ref{eq:HoverMhorizon}) are not in conflict, since the former is evaluated for a fixed $k$, and the latter is evaluated for different $k$-modes at horizon crossing.}
\begin{equation}
\label{eq:HoverMhorizon}
\left({H \over M}\right)_{k = a H} \propto k^{- \epsilon}
\end{equation}
to write the power spectrum modulation as:
\begin{equation}
{\delta P \over P}\left(k\right) = 1 +  \beta_0 \left({H \over M}\right)_{k_* = a H} \left({k \over k_*}\right)^{- \epsilon}.
\end{equation}
Note that we now have a logarithmic dependence of the modulation on $k$, and the amplitude of the modulation {\em decreases} at short wavelengths. We can compare this to the NPH solution (\ref{eq:PNPH}) by noting that the parameter $y_c$ depends on $k$ as \cite{Easther:2002xe}
\begin{equation}
y_c \equiv \left({k \over a H}\right)_{k = a M}  \propto k^{\epsilon}.
\end{equation} 
We can then write the NPH power spectrum modulation (\ref{eq:PNPH}) as:
\begin{equation}
{\delta P \over P}\left(k\right) \simeq 1 +  {1 \over 2} \left({H \over M}\right)_{k_* = a M} \left({k \over k_*}\right)^{- \epsilon} \sin\left[2 y_c\left(k\right)\right].
\end{equation}
We then see that the constraining the EFT modulation to be independent of the choice of initial hypersurface gives a modulation amplitude of the same order as that of the NPH formalism, and with the same logarithmic $k$-dependence. However, we do not recover the periodicity of the modulation using this rough analysis. It would be interesting to perform a more sophisticated analysis incorporating hypersurface-independence into the full EFT formalism to determine the detailed form of the modulation. A self-consistent EFT description incorporating with this property is, however, likely to be problematic, since it will necessarily involve the introduction of a second physical scale $H$ into the EFT expansion \cite{KSJPvDSprivate}. 

\section{Conclusions}
\label{sec:conclusion}

The purpose of this paper has been to carefully compare the astrophysical consequences of the NPH and BEFT approaches to calculating the trans-Planckian corrections to the power spectrum of primordial perturbations.   The observability of NPH-induced corrections to the power spectrum was carefully examined by us in \cite{Easther:2004vq}, and also in \cite{Okamoto:2003wk}. That analysis focused on the CMB, but since the correction to the spectrum is a slowly changing function of the comoving wavenumber $k$, including high quality information from Large Scale Structure surveys or the Lyman-$\alpha$ forest will improve the chances of detecting a trans-Planckian modulation, but will not introduce any qualitatively new features. 

The BEFT formalism predicts that the amplitude of the modulation increases linearly with $k$, and thus it can be undetectably small at some scales of astrophysical interest, but induce a correction of order unity at wavelengths two orders of magnitude smaller.  If this breakdown occurs in the range of $k$ values that source structure at cosmological scales, then we will can expect to see  a large feature in the CMB spectrum at the $k$ value where the breakdown occurs, and the spectrum at shorter scales can only be computed via a full quantum gravity / stringy calculation. In the NPH approach, we expected a small but hopefully observable correction to the standard inflationary expectation for the primordial spectrum.  In the BEFT approach, the spectrum is either effectively unmodified or will be dominated by new physics.

Our analysis does not break new ground in the BEFT approach itself, but rather carefully explores the consequences of the spectrum derived by Greene, Schalm, Shiu and Van der Schaar in \cite{Greene:2005aj}.  The cosmological consequences are potentially extreme, in that the corrections to the spectrum can be of the same order as the spectrum itself. 
In the light of our apparently startling conclusions it is worth reflecting on the overall applicability of the BEFT approach within cosmological spacetimes. 
The key conceptual problem is that there are two mutually incompatible defintions of  ``large scale''   in this problem.  From the BEFT side,  we expect the approach to break down at high momenta -- which is to say at short wavelengths or large values of $k$, as is the case in any field theory with a cut-off.  From the cosmological perspective, however, the controlling parameter is $H/M$, and this parameter (slowly) decreases as the universe inflates, leading us to expect a greater contribution from high energy effects on very long wavelength perturbations -- as is seen in the NPH case. Our application of the power spectrum modulation calculated in Ref. \cite{Greene:2005aj} over many decades in $k$ is in this sense too literal: cosmology forces us to extend the formalism into regimes where it is known to break down. Given {\em any} choice of boundary hypersurface, only two decades of modes with wavelength longer than the UV cutoff will reside within the horizon. It is not at all clear how to self-consistently treat modes outside this comparatively narrow range of scales. 
With this in mind, we looked more carefully at the freedom that exists inside the BEFT formalism for ``running'' in the parameter $\beta$ to restore some of the time invariance of the de Sitter case that is broken when BEFT effects are incorporated. We make a crude phenomenological analysis of this issue, showing that an appropriately running coupling does  in fact restore some of the time invariance and that when this is done, the BEFT and NPH approaches coincide far more closely.  As noted in Section \ref{sec:runningbeta}, resolving this problem will require creating effective field theory methods that remain self-consistent even in an expanding universe.   As it stands, however, our analysis highlights the risks implicit in using the BEFT spectrum of equation~(\ref{eq:EFTmod}) over a large range of $k$ values. Despite the considerable theoretical appeal of the BEFT formalism, the self-consistent BEFT calculation of trans-Planckian modifications  to  the  full cosmological perturbation spectrum remains an open question.

\section*{Acknowledgments }
We are grateful to Christophe Ringeval for advice on modifying CAMB to remove $\ell$-space interpolation in order to produce the CMB spectra plotted in this paper. We thank Amjad Ashoorioon, Robert Brandenberger, Scott Burles, and Eugene Lim for a number of discussions.  We are particularly grateful to Brian Greene, Koenraad Schalm, Gary Shiu and Jan Pieter van der Schaar for many fruitful discussions about the BEFT formalism,  and their insights have contributed significantly to this paper.   RE  is
supported in part by the United States Department of Energy, grant
DE-FG02-92ER-40704.  HVP is supported by NASA through Hubble Fellowship
grant \#HF-01177.01-A awarded by the Space Telescope Science
Institute, which is operated by the Association of Universities for
Research in Astronomy, Inc., for NASA, under contract NAS 5-26555.

\end{document}